# Looking for grass-root sources of systemic risk: the case of "cheques-as-collateral" network

Michalis Vafopoulos
Aristotle University of Thessaloniki
vafopoulos.org

**ABSTRACT**
The global financial system has become highly connected and complex. Has been proven in practice that existing models, measures and reports of financial risk fail to capture some important systemic dimensions. Only lately, advisory boards have been established in high level and regulations are directly targeted to systemic risk. In the same direction, a growing number of researchers employ network analysis to model systemic risk in financial networks. Current approaches are concentrated on interbank payment network flows in national and international level. This work builds on existing approaches to account for systemic risk assessment in micro level. Particularly, we introduce the analysis of intra-bank financial risk interconnections, by examining the real case of "cheques-as-collateral" network for a major Greek bank. Our model offers useful information about the negative spillovers of disruption to a financial entity in a bank's lending network and could complement existing credit scoring models that account only for idiosyncratic customer's financial profile. Most importantly, the proposed methodology can be employed in many segments of the entire financial system, providing a useful tool in the hands of regulatory authorities in assessing more accurate estimates of systemic risk.

## INTRODUCTION

The global financial crisis that started in 2007 epitomized the role of strong financial ties as a carrier for propagation of shocks. The rapid viral spreading of the financial turmoil from the US sub-prime mortgage market to mammoth international financial institutions manifested that global economy is based on numerous, strong, unexplored and unanticipated interdependencies among networks in credit, trade, investment and supply chains. Consequently, it is needed an innovative methodology that models the systemic risks of financial networks and can be used to design effective policies to reduce conflicts between local and global interests [1].
In this context, the European Central Bank studies the emerging aspects of systemic risk [2] and the European Commission established the European Systemic Risk Board as an independent EU body to anticipate systemic risks within the financial system. At the same time, a growing number of studies and conferences examine various aspects of financial networks and specifically, the modeling of systemic risk using mathematical network analysis [3]. Current analysis is focused on interbank payment network flows in national and international level [4].

In the banking sector analysis, existing credit scoring models account only for idiosyncratic customer's financial profile and do not anticipate systemic risk factors (e.g. a potential domino effect caused by the bankruptcy of a central node-customer) [5]. Financial volatility [6] and related statistical modes (e.g. VaR [7]) are not anymore adequate approximations for risk factors. Banks have no information about the negative spillovers of disruption to a financial entity in their lending network because they are not directly account for linkages among customers in order to determine counterparty losses, predict and anticipate failure cascades. Consequently, their financial statements and reports to regulatory authorities do not account for this serious source of systemic risks in the financial system.

In our research, we extend the current analysis of interbank networks to intra-bank financial risk interconnections, namely the "cheques-as-collateral" network. The proposed model address and quantifies systemic default risk factors like chain bankruptcy caused by central customers of the "cheques-as-collateral" network. Based on the experimentation with a real dataset of 422 funded customers and using as collateral 783 cheques, we found out that an initial failure of five customers, representing the 17% of the total value, results the subsequent failure of 15 customers, which represent the 41% of network's total value.

According to [8]: *"If the recent crisis has taught us anything, it's that risk to our system can come from almost any quarter. We must be able to look in every corner and across the horizon for dangers. During the crisis, our system was not able to do this. Financial systems are evolving rapidly and this evolution is endogenous"*.

The paper is organized as follows. The first Section underlines the growing importance of networks in modeling complex systems. Section 2 briefly discusses the main definitions of systemic risk and Section 3 states the scope of the paper. Relevant literature is reviewed in the next section. Section 5 introduces the "cheques-as-collateral" network. The next section refers to the dataset that is used to estimate the proposed model. Section 7 analyzes the model and presents the results. The final section concludes and discusses ideas for further research.

1. **The growing importance of networks**

Barabasi's work on modeling Internet and the Web networks revealed their self-similar structure [9], [10]. The power-law degree distribution, together with the "small-world" property of networks [11], initiated a vast amount of research on the statistical properties and implications of real networks in general [12], [13]. Network theory contributes useful methodologies developed in other disciplines such as Web analysis, epidemics, sociology and mathematics to enable risk assessment. For instance, the identification of hubs and authorities [14] and the diffusion of information in the Web [15] help to identify the driving forces of financial contagion in the banking sector. Financial networks are difficult to model because they are formed by dynamic and strong strategic interactions in different levels involving heterogeneous players.

Nowadays, in the financial industries, stress testing and a more detailed analysis of the risk in the micro level (i.e. loan-level) rather than aggregate data, have replaced correlation risk models [8]. The focus is *"Where possible, it is useful to do analysis at the*

*individual loan level. And because it is difficult to reach agreement upon an analytical approach and the pace of adoption is slow, it may be best to start with simple risk measurements and work toward more sophisticated methods."*

In this line of research, a growing number of researchers examine various aspects of financial networks and specifically, the modeling of systemic risk using mathematical network analysis [3]. Current analysis is focused on the systemic risk of intra-bank payment network flows in national and international level.

## 2. Systemic risk

Various definitions (often complementary) for systemic risk exist in the literature [16] [3] and is beyond the scope of the present paper to analyze them.

A large price movement in the market or the announcement of company losses is not characterized as systemic risk. Systemic risk is not related to individual events but to collective reactions on them. The failure of large and interconnected financial institutions could cause negative externalities to economy as a whole. *"Since the costs of a failure do not fall exclusively on the failing institution, there is an incentive for firms to take excessive risk and to invest less in risk management than is socially optimal"* [3]. An analogous argument is valid for individual bank customers. Let us briefly describe systemic risk as:

- The risk of disruption to a financial entity with spillovers to the real economy.
- The risk of a crisis that stresses key intermediation markets and leads to their breakdown, which impacts the broader economy and requires government intervention.
- Adverse general equilibrium amplification and persistence.

According to the network perspective, systemic risk is:

- The risk that critical nodes of a financial network cease to function as designed, disrupting linkages.
- Financial contracts with externalities.

Systemic risk is an issue of primer importance for the stability of the banking system and the financial contagion capability of particular banks contributes a major part in this risk. Only recently, systemic risk measures have been developed to account for the systemic risk in the bank sector [17].

## 3. Scope of the paper

Risk is transferred, not transformed. This is true not only among banks or other financial entities but also among bank customers. Current risk systems do not account for linkages and failure cascades and are not able to determine counterparty losses. The mass *aggregation* of financial data (e.g. bank loans in financial statements, synthetic derivatives) eliminates valuable information about the origins of systemic risk.

Our work initiates the discussion of exploring the systemic risk of individual bank customers to the bank's loan portfolio. In particular, we are focused on the following tasks:

- Collect data on interconnected bank customers and identify the network structure.
- Assess the risk to a bank from its customer's position in the "cheques-as-collateral" network.
- Identify "systemically important" nodes whose collapse can significantly impact

- the network.
- Develop the ability to measure and rank nodes by their fragility.
- Apply the growing understanding of network dynamics to intentionally design robust customer networks.

The issue of analyzing a new dimension of the systemic risk and the associated contagion process for an individual bank is addressed. This task is undertaken by creating a network model for estimating the distribution of losses for a bank caused by the failure of individual customers, which have been funded by the bank. The followed approach is inspired by the Systemic Risk Network Model that accounts for inter-bank failures [17].

## 4. Relevant literature

Literature about systemic risk and contagion is fairly new and limited. This section is by no means an exhaustive review of the fast evolving literature in the network analysis of systemic risk, but only highlights some major developments in relative research efforts. [18] created a theoretical framework to investigate interbank lending and systemic risk and concluded that interbank lending could be used for prudential control. [19] simulated the effects of failure in major US banks. Based on his work many scholars investigated financial contagion in different countries (e.g. UK [20]).

Networks have been applied in several fields of economics (for reviews see [21], [22]). The main application of network theory to financial contagion refers to modeling interbank markets based on data from banks' balance sheets and interbank payments [23]. Banks are the nodes, inter-connected if financial flows and exposures exist among them. [24] investigate the effect of variations in different parameters (e.g. the bank's capitalization) on bank failures due to contagion. [25] develop a model of contagion in arbitrary financial networks and conclude that financial systems may exhibit a robust-yet-fragile tendency. Similar models have been applied to national interbank markets. For instance, [26] provide a time-varying analysis of the Italian overnight market network and [27] assess the potential for contagion in the Swiss interbank market.

## 5. The "cheques-as-collateral" network
### 5.1. Introduction

Unlike the rest modern economies, it is a common practice in Greek B2B and B2C commerce to make the payments of liabilities by cheques with late maturity. Many professionals put these cheques as collateral to receive working capital credit from commercial banks. The approved credited amount as percentage of the cheques' value varies from 70% to 90% for more trustworthy customers. Bank receives commission for the provided services raging approximately to the 7‰ of cheque's amount plus the interest charge on capital (today varies between 8%-10%). According to the inter bank analysis agency, Tiresias (teiresias.gr), the total value of bounced cheques was more than 3 billion euros in 2009, 2.3 billions in 2010 and 1.4 billion during the first six months of 2011.

If a cheque issuer, bankrupts then cheques that has issued cannot be paid (often called "bounced cheques"). As results of this, the bank customer, who has used her cheques as collateral has to undertake the failed payment. In such case, the risk for the lending bank is similar to the default loan risk. If a cheque recipient, who has received credit based on her customers' cheques as collateral, bankrupts, it is assumed that she is not able to pay her cheques, but her customers' cheques will be settled. In this case, the risk for the lending bank comes from the possibility that the defaulted customers' cheques become bounced. This kind of risk is less important than the default loan risk because cheque issuers have not been bankrupted.

### 5.2. Assumptions

For our model, bank customers are considered to be cheque issuers or/and recipients. Check recipients can use their incoming cheques, which may have different maturity dates (commonly received as payments by their customers in exchange to good or service provision) as collateral to working capital credit. The "cheques-as-collateral" network is formed as follows: if customer $i$ have issued one or more cheques to customer $j$, the link $i \xrightarrow{w} j$ exists with weight w. W is equal to the fraction of the value of cheques that customer $i$ have issued to customer $j$, $d_{ij}$, to the total value of cheques in euros, $V$, received by the bank as collateral to loans during a specific period of time (model variables are presented in Table 1). When a customer bankrupts, it is assumed that her issued cheques (out-degree) are not paid, but the received cheques (in-degree) are fully paid by the issuers to the bank, which has received them as collateral from the failed customer.

It is also assumed that all cheques have been issued in the beginning of the year and they will be paid at their maturity date. When a cheque recipient uses a new cheque as collateral in order to receive funding, the weight of the new cheque is added in the existing weight between cheque recipient and cheque issuer.

| | |
|---|---|
| n | total number of bank customers |
| i | cheque issuer, 1, 2, …, p |
| j | cheque recipient, 1, 2, …, m |
| $l_i$ | loss to the network given failure of customer $i$ |
| $d_{ij}$ | total exposure of customer $i$ to customer $j$ or total value of cheques in euros issued/paid by customer $i$ to customer $j$ during a specific period of time |
| $V$ | total value of cheques in euros received by the bank as collateral to loans during a specific period of time |
| k | stage of financial contagion caused by an initial shock |
| $u_i^k$ | failure/bankruptcy threshold for customer $i$ at stage k (0,1] |
| c | percentage of total losses that collapse the node (0,1] |
| $w_{ij}$ | link weight, $\frac{d_{ij}}{c}$, (0,1] |
| $i \xrightarrow{w} j$ | customer $i$ have issued one or more cheques to customer $j$, with weight w |
| θ | a state variable that indicates whether a bank customer is failed or not {0,1} |
| $D^k$ | set of failed customers at stage $k$, [0,422] |

Table 1: description of the "cheques-as-collateral" model variables

## 6. Data

Our sample is the "cheques-as-collateral" network of a leading Greek bank's local branch data collected from 31/03/2009 until 31/03/2010. The network is based on the 33 customers of this branch that have been financed by setting as collateral their received cheques. 422 nodes that are connected through 450 links form the network under consideration (Table 2). Each link is weighted by the fraction of total exposure of customer *i* to customer *j*, to the total value of issued cheques. For example, an issued cheque of 10,000 euros weights 0.232 (=100,000/4,306,735). The examined network is sparse and not weakly connected because it contains unconnected components. The "cheques-as-collateral" network is depicted in Figure 1 based on [28] and the darker link lines indicate higher weight for the link between two customers. The examined sample network is considered to be scale-free, as many other real networks (e.g. the Web [10]) with power-law exponent of the in-degree distribution to equal 1.3).

| | |
|---|---|
| *Nodes (customers)* | **422** |
| *Links (issued cheques)* | **450** |
| *Average degree* | **2,13** |
| *Max in-degree* | **108** |
| *Max out-degree* | **3** |
| *Weakly connected* | **False** |
| *Average path length* | **1.72** |
| *Diameter* | **6** |
| *Power law exponent* | **1.3** |
| *Total value of issued cheques (€)* | **4,306,735** |
| *Funded customers* | **33** |
| *Number of Communities* | **33** |

Table 2: descriptive statistics for the "cheques-as-collateral" network

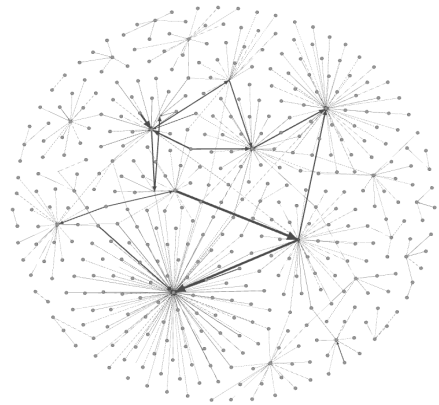

Figure 1: visualization of the "cheques-as-collateral"

Many real networks are characterized by their modular structure. Specific groups of nodes are more densely interconnected with a group of nodes than with the rest of the network (also called communities or cohesive subgroups or clusters [29]). The network is consisted of 33 distinct communities formed by the funded customers (Table 2).

| Node | In-degree | Weighted in-degree |
|---|---|---|
| 1002 | 108 | 28.64 |
| 1001 | 55 | 10.57 |
| 1011 | 41 | 15.47 |
| 1019 | 35 | 10.69 |
| 1013 | 23 | 3.29 |

Table 3: the top five in-degree customers and their associated weighted in-degree

The node in-degree in the "cheques-as-collateral" network describes the number of different cheque issuers that a funded customer uses as collateral. Weighted in-degree refers to the value of cheques that a funded customer contributes in the "cheques-as-collateral" network. In the examined network, the top funded customer brings also the largest number of cheques (Table 3). 1002 belongs to the house construction industry and has a long term and trustworthy credit history. 1001 is specialized in supplying products for personal Hygiene & Sanitation and Equipment Office & Stores. 1011 shells plastic tanks and pipes, while 1029 is the biggest gas-petrol station in the area.

| Node | Betweenness centrality |
|------|------------------------|
| 1029 | 150.5 |
| 1011 | 109.0 |
| 1031 | 80.5 |
| 1017 | 74.5 |
| 1019 | 68.5 |

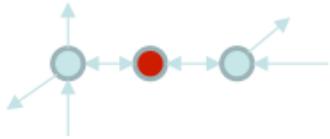

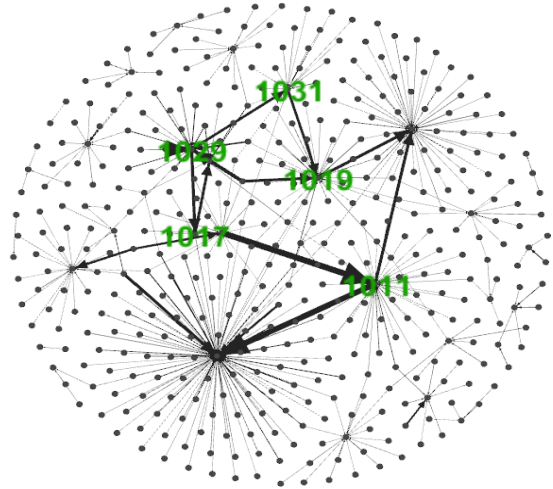

Table 4: Top five customers with respect to betweenness centrality

Figure 2: top five customers with respect to betweenness centrality

Node *betweenness centrality* is the number of shortest paths going through the node and describes the extent to which a node lies between other nodes in the network. This measure takes into account the connectivity of the node's neighbors, giving a higher value for nodes which bridge clusters, indicating how often a node is found on a shortest path between two nodes in the network. For a graph $G := (V, E)$ with $n$ nodes, the betweenness $C_B(v)$ for node $v$ is given by:

$$C_B(v) = \sum_{s \neq v \neq t \neq V} \frac{\sigma_{st}(v)}{\sigma_{st}}$$

where $\sigma_{st}$ is the number of shortest paths from $s$ to $t$, and $\sigma_{st}(v)$ is the number of shortest paths from $s$ to $t$ that pass through a node $v$.

In the "cheques-as-collateral" network betweenness centrality captures the connectivity among the cheque issuers that are used as collateral for funded customers. It could be an indication of how often funded customers exploit the same cheque issuers, and thus increasing the systemic risk in the network. In the network under investigation, some of customers with high funding are in the top positions because use cheque issuers with high involvement in other funded customers (Table 4 and Figure 2).

# 7. The model

The most interesting case of financial contagion in the "cheques-as-collateral" network emerges when a check issuer cannot pay her cheques and as result of this, the recipients of her cheques cannot also pay (part or all) of their issued cheques. Current risk management models are not capturing this domino effect, because are based on idiosyncratic financial evaluation for each customer. Often existing approaches evaluate customers in a binary format (good or bad), according their repayment performance over a fixed time period (see for instance [30]). The two mainstream approaches occupied in credit scoring are discriminant analysis (see for instance [31]) and logistic regression analysis (see for instance [32]). The majority of these methodologies are time invariant and rank bank customers mainly according to their characteristics as have been filled on their application form and their credit history. Banks are not accounting for linkages among customers and thus cannot determine *counterparty losses*, predict and anticipate *failure cascades*. The bank has no information about the negative spillovers of disruption to a financial entity in its lending network.

Current analysis aims to complement existing evaluation procedures by offering additional information with respect to the systemic effect of every customer.

For our model, $F$ denotes the set of failed bank customers in the initial shock, and $L(F)$ are the *losses* to the system if "scenario F occurs". $L$ includes the decrease in total value of cheques received by the bank as collateral to loans that has been caused due to losses of the customers, which fail in the initial shock, and the losses due to the contagion generated by these customers. Following the analysis of [17], since we model contagion as a deterministic process, the customers that fail due to contagion of initially failed customers in $F$, are unique. If $C(F)$ is defined to be the set of customers that fail due to contagion whose source is F, then the *total loss* for a given scenario F is:

$$L(F) = \sum_{i \in F} l_i + \sum_{i \in C(F)} l_i$$

If it is assumed that the failure probabilities of customers are independent, probability of occurrence of this loss is given by:

$$P(F) = \prod_{i \in F} p_i \prod_{i \in F^\sim} (1 - p_i)$$

where $F^\sim$ is the complement of $F$ (the not failed bank customers).

Given the above definitions, we introduce a discrete model of financial contagion for assessing systemic risk in the "cheques-as-collateral" network. First, let us assume that at every stage of contagion and for each customer $j$, there is a specific "threshold" $u_j^k$, such that if a customer's exposure to previously defaulted customers exceeds the threshold, that customer will also fail. $D^k$ is the set of all customers that have failed by stage $k$. Hence, customer $j$ will fail at stage $k + 1$ if the following condition holds:

If $\sum_{i \in D^k} w_{ij} \geq u_j^{k+1}$ then j defaults: $j \in D^{k+1}$    (1)

where $\sum_{i \in D^k} w_{ij}$ is the sum of defaulted exposures of customer $i$ to $j$ at stage $k$ and $c$ is defined to be the percentage of total losses from unpaid cheques that drive every bank customer to failure (i.e. the "failure or bankruptcy threshold"). The failure threshold for customer $j$ at stage $k$ is given by the formula $u_j = c \sum_{i=1}^{n} w_{ij}$.

It is also necessary to define a state variable in order to indicate whether a customer is failed or not at stage $k$ of the contagion process as:

$$\theta_j^k = \begin{cases} 1 \text{ if } \sum_{i \in D^{k-1}} w_{ij} \geq u_j^k \\ 0 \quad\quad\quad\quad\quad\quad\quad else \end{cases} \quad (2)$$

As an illustrative example, consider a small part of the "cheques-as-collateral" network consisting of six customers and six issued cheques and weights as depicted in Figure 3. Let us assume that the percentage of total loss from unpaid cheques, which drive every bank customer to failure $c$, equals 0.5, that is, at least, 50% of the value of the received cheques has not been paid.

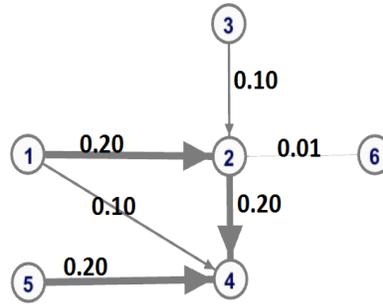

Figure 3: an example of "cheques-as-collateral" network consisting of six customers and six issued cheques with the associated weights.

It is also assumed that at stage $k = 0$ customer 1 fails to pay the cheques issued by her. The failure thresholds for the recipients of her cheques, $u_2$ and $u_4$ equal 0.16 and 0.25, respectively. The sum of the defaulted exposures for customer 2 exceeds her failure threshold and according to formula (1) also fails at stage 1. Contrastingly, customer 4 survives in this stage because her defaulted exposures are less than the corresponding threshold. In the following stage, due to the financial contagion customer 4 is affected by the failure of 2 and also fails.

### 7.1. Algorithm
The algorithm that is used to calculate the propagation of financial contagion in the "cheques-as-collateral" is minutely presented in the following lines[1].

---

[1] Matlab code and data are available upon request.

**Step 0**
1. Assume a set of criteria for the failure of every customer ($c$). Here it is assumed that $c$ is the percentage of the total amount of unpaid cheques that drives every bank customer to failure.
2. For a given "cheques-as-collateral" network, calculate the weighted adjacency matrix ($W$).
3. Calculate the failure threshold for every customer $j$, $u_j = c \sum_{i=1}^{n} w_{ij}$. It is assumed that this threshold remains constant at every stage $k$.
4. Assume a set of customers that initially fail to pay their cheques ($D^{k=0}$). This set can be chosen by some relevant criterion. In our case, five customers with the highest weighted out-degree have been selected to collapse at stage $k = 0$ and their relevant state variable $\theta$ is set to be equal to 1.

**Step 1**
1. Calculate the sum of the defaulted exposures of failed customer $i$ to $j$, $\sum_{i \in D^k} w_{ij}$.
2. Compare the calculated defaulted exposure failure threshold of customer $j$. If (1) holds then customer $j$ fails at this stage, $j \in D^{k=1}$.
3. Update $D^k$ with the failed customers.

**Step 2**
1. Repeat Step 1 until $D^k = D^{k+1}$.

The first assumption in the initial step can be relaxed to include more sophisticated and personalized criteria for customer bankruptcy. Similarly, the fourth assumption in this step can be modified to encapsulate credit risk (e.g. customers form sectors with high risk and interconnections) and other network criteria (e.g. eigenvector centrality) to select the initially failed set of customers.

### 7.2. Results
The "cheques-as-collateral" network under investigation is formed by 422 customers and 783 cheques, which are aggregated to 450 financial links (Figure 4a). Let us assume again that the percentage of total losses from unpaid cheques that drive every bank customer to failure $c$, equals 0.5, that is, at least, 50% of the value of the received cheques has not been paid. It is also assumed that at stage $k = 0$, five customers with the highest weighted out-degree fail to pay their cheques, causing a 17.11% decrease in the total value of the cheques network. Applying algorithm (1), results the failure of 4 customers at stage 2, increasing the devaluation to 26.62% (Figure 4b). The financial contagion expands further to four stages resulting the failure of 15 customers, which represent the 40.68% of underlying cheques total value (Figure 4f).

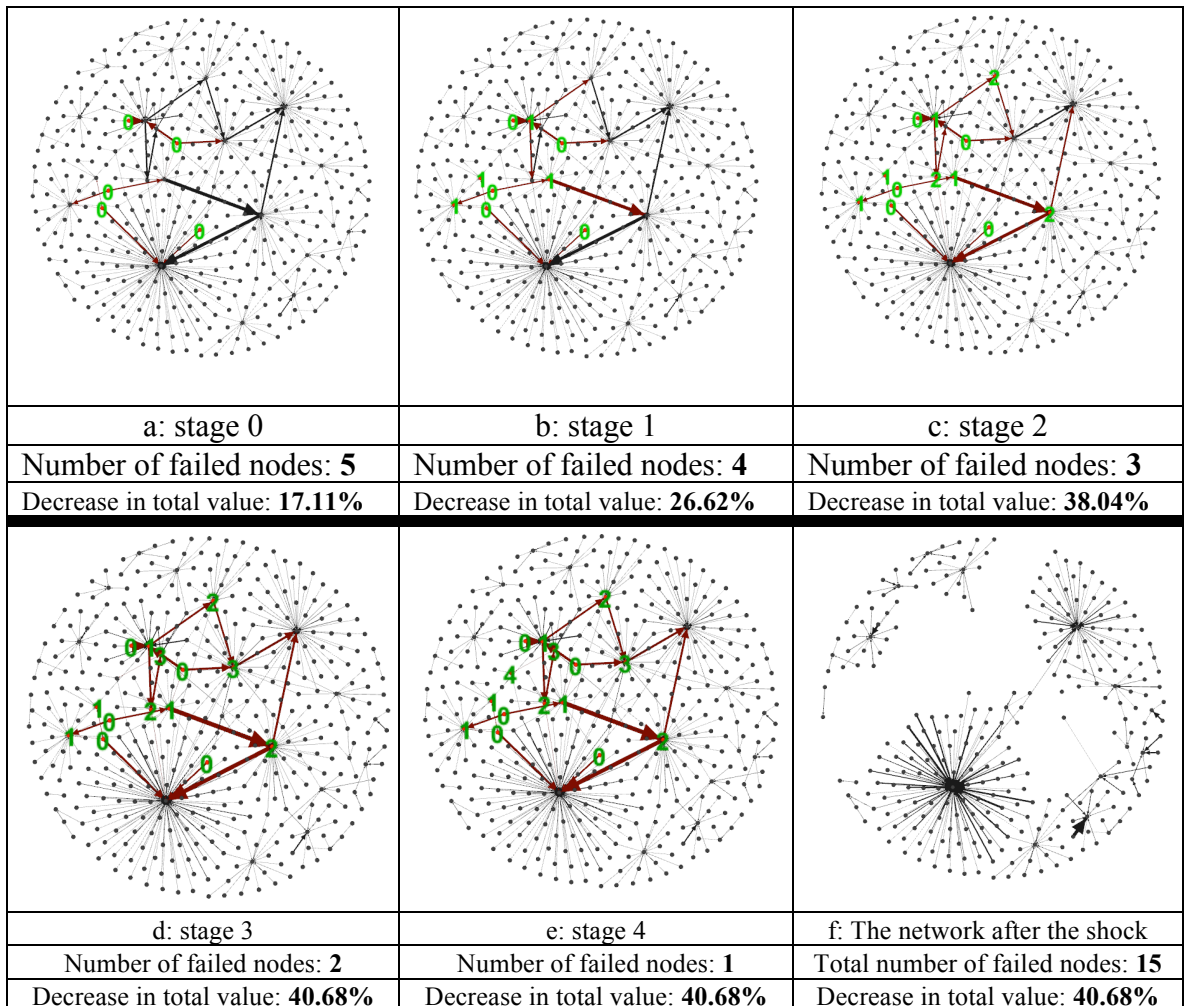

| a: stage 0 | b: stage 1 | c: stage 2 |
| --- | --- | --- |
| Number of failed nodes: **5** | Number of failed nodes: **4** | Number of failed nodes: **3** |
| Decrease in total value: **17.11%** | Decrease in total value: **26.62%** | Decrease in total value: **38.04%** |
| d: stage 3 | e: stage 4 | f: The network after the shock |
| Number of failed nodes: **2** | Number of failed nodes: **1** | Total number of failed nodes: **15** |
| Decrease in total value: **40.68%** | Decrease in total value: **40.68%** | Decrease in total value: **40.68%** |

Figure 4: financial contagion in the "cheques-as-collateral" network at every stage (green numbers indicate the nodes that collapse in the every stage).

### 7.3. Evaluating the systemic risk of a bank customer
#### 7.3.1. Case 1: stage-by-stage loss

For every Greek commercial bank it is important to make a decision about additional financing of existing customers by accepting new cheques as collateral. In this context, a systemic risk measurement for each customer is proposed. The measurement is given by applying the algorithm in Section 7.1 in the case of failure of an individual node and by calculating the resulting financial contagion (i.e. the total number of failed nodes and the total value decrease). Table 5 presents the case of the five most "dangerous" customers in terms of causing financial contagion. If it is assumed that the percentage of total losses that drive every customer to failure, $c$, equals 0.5, then 1005 causes the most serious domino effect in the network. In this case, total loss has calculated by assuming that loss in every stage is *uniformly* evaluated. It is noteworthy that the biggest customer (1002), who has been used the highest number of cheques as collateral (108, while the second has 55) and represents almost the 30% of the total of amount of cheques, is not in the list with the most "dangerous" customers. This happens because the proposed evaluation criterion accounts for the risk arising from the locus of customers in the network.

The uniform loss hypothesis could be relaxed in accordance to different credit scoring characteristics and goals. For instance, a credit risk policy may prioritize the anticipation of financial contagion in early stages or the aversion to specific sectors of the economy and the diversification of the cheques received as collateral. We examine the first scenario of penalizing loss by the half penalty of the preceding stage:

$$Total\ loss = \sum_0^k (Loss\ at\ stage\ k) \frac{1}{2^{k+1}} \qquad (3)$$

In Table 6 the total adjusted losses is calculated by weighting stage 0 loss with 0.5, stage 1 losses with 0.25 and stage 2 losses with 0.125. This new criterion of systemic risk overturns the ordering of customers of the examined network (e.g. 1011 is relegated to the second position of the most "dangerous" customers).

| Nodes | Stage 0 | | Stage 1 | | Stage 2 | | Total loss | Total adjusted loss |
|---|---|---|---|---|---|---|---|---|
| | Failed nodes (number) | Value decrease (weight) | Failed nodes (number) | Value decrease (weight) | Failed nodes (number) | Value decrease (weight) | | |
| 1005 | 1 | 5.11 | 1 | 7.15 | | | 12.26 | 4.34 |
| 1029 | 1 | 4.41 | 2 | 4.27 | 1 | 0.21 | 8.89 | 3.29 |
| 1011 | 1 | 7.15 | | | | | 7.15 | **3.57** |

Table 6: Total uniform loss and total adjusted loss due to the failure of an individual node.

### 7.3.2. Case 2: composite loss

As it was shown by our analysis, systemic risk exists in the "cheques-as-collateral" network. The stage-by-stage loss function (3) could become a more accurate proxy of the potential damage caused by a single customer by taking into account her weight in the network. According to the composite loss hypothesis, each customer receives a score depending on her weight in the network, the $l_j$ potential damage will cause to the network if she bankrupts and the average weight of each funding cheque with respect to the failure threshold $u_j$. The composite loss of each customer is given by the following formula:

$$g_j = \left(\frac{\sum_{i=1}^{422} W_{ij}}{d^{in}} - u_j\right) + l_j + \sum_{i=1}^{422} W_{ij} \qquad (4)$$

The five customers with the highest composite loss index ($g$) are presented in Table 7.

| Customer | Composite loss (g) | Weight |
|---|---|---|
| 1029 | 15,78 | 12,54 |
| 1002 | 14,58 | 28,64 |
| 1005 | 13,88 | 2,91 |
| 1011 | 11,51 | 8,32 |
| 1019 | 6,80 | 8,26 |

Table 7: Customers with the highest composite loss index and their underlying weight.

Customers with low weight could be characterized by high composite loss index (e.g. 1005) and oppositively, customers with high weight are described by low composite loss (e.g. 1002, 1001). It seems that the composite loss index captures systemic risk factors, apart from the weight of the customer.

### 7.3.3. Case 3: systemic risk assessment

The composite loss function could be extended to account also for the state of cheque issuers in order to become a more accurate tool in the decision-making process for extra funding in existing customers. Specifically, we consider the "other side" (i.e. cheque issuer) of the systemic risk caused by the fact that customer $j$ intents to use as collateral a cheque issued by customer $i$. The "systemic risk assesment" of issuer $i$ for its new cheque that is brought as collateral in the network is given by:

$$r_i^k = \sum_{j=1}^{j=n} \left[ \left( \frac{w_{ij}}{u_j} \times g_j \right) + r_j^{k-1}|_{\alpha v\ w_{ij} \neq 0} \right] \qquad (5)$$

In such case, banks can put upper limits not only for funded customers, but also for cheque issuers in order to minimize their exposure to financial contagion (five customers with the highest weighted out-degree are presented in Table 8).

| Node | Systemic risk ($r_i$) | Weighted out-degree |
|---|---|---|
| 216 | 18,63 | 3,60 |
| 128 | 12,09 | 5,33 |
| 127 | 9,06 | 3,60 |
| 029 | 3,07 | 2,84 |
| 067 | 1,78 | 1,76 |

Table 8: Customers with the highest systemic risk and their associative weighted out-degree.

Despite the fact that 216 and 127 share the same weighted out-degree, the latter is characterized by substantially lower systemic risk factor. That happens because 216 is connected to 1005 and 1013, which are characterized by high weighted out-degree but low failure thresholds. On the contrary, 127 is connected to 1029 which has higher failure threshold. Thus, 127 enjoys the fact that 1029 is less sensitive to financial contagion.

## 8. Discussion and further research

Globalized practices transformed financial transactions into a highly interconnected and complex system. Even the best analysts find hard to calculate the associated risk of every transaction without missing some important factor and underlying relation. Existing models of financial risk have been build on the dogma "too big to fail" and do not account for the risk of disruption to a financial entity and its spillovers to the real economy. Most methodologies fail to create accurate estimates of systemic risk factors because are based on *aggregated data* of financial transactions. Only recently, researchers and regulating bodies have been started to focus on the risk that emerges when critical nodes of a network cease to function as designed, disrupting linkages.

In the present paper, a new model for estimating financial contagion in micro-level is proposed. Specifically, we analyze the lowest-level dimension of systemic risk and the associated contagion process for an individual bank by creating a network model to estimate the propagation of losses caused because of failure of some funded customers.

Based on the experimentation with real data, we found that an initial failure in the 17% of the total value of cheques, results the subsequent failure of the 41%.

The employment of current analysis could be twofold: (a) by complementing traditional credit scoring models with additional information about the systemic risk of individual customers and (b) by providing a useful tool for regulatory authorities in assessing more accurate estimates of systemic risk in many segments of the real economy. Specifically, the proposed model can be applied and enriched with additional financial transactions and entities (e.g. loans, invoices) in order to capture grass-root sources of risk. For instance, building and monitoring the network of private debt in micro-level could offer a more efficient tool in anticipating financial contagion for national interbank analysis agencies and international regulators.

The proposed approach should be further explored with simulated and real datasets in order to identify the full spectrum of its statistical characteristics. Business implications in the financial sector should be also addressed. For instance, specific complementarities with existing risk models and efficiency measures for financial institutions [33] should be analyzed in detail.